# PERTURBATION THEORY CONFRONTS OBSERVATIONS : IMPLICATIONS FOR THE "INITIAL" CONDITIONS AND $\Omega$

F. R. BOUCHET[1], R. JUSZKIEWICZ[1,2]

[1] *Institut d'Astrophysique de Paris, CNRS, 98bis Boulevard Arago, F–75014 Paris France.*
[2] *Copernicus Astronomical Center, ul. Bartycka 18, Warszawa, Poland.*

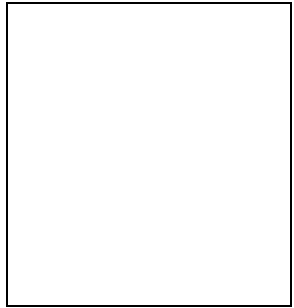

**Abstract**

This paper covers the material of our two talks. We describe a series of projects based upon perturbative expansions to follow the gravitational evolution of the one point probability distribution functions (PDFs) for the density contrast field and for the divergence of the corresponding velocity field. The Edgeworth expansion greatly simplifies the problem. Indeed, the PDF can be described through several low order moments, with the first non-trivial contribution coming from the skewness, or third moment.

Gaussian initial conditions imply that the skewness of the smoothed density field is proportional to the square of the variance. The constant of proportionality is then a function of the logarithmic slope of the variance versus scale, and is otherwise independent of scale. This constant is also practically independent of the density parameter $\Omega$, and has very nearly the same value when distances are estimated by redshifts. To show the latter property, we also briefly discuss our Lagrangian approach to perturbation theory. Finally, the observed skewness does depend, of course, on the relation between the distribution of light and mass; we describe the case of local biasing.

For the divergence of the velocity field, the skewness of its PDF is also proportional to the square of its variance, but the ratio now strongly depends on the value of $\Omega$. This offers a new method to determine the density parameter which does not involve comparisons with the density field; it should be independent of assumptions concerning the way light traces mass.

Comparisons with numerical simulations show that these theoretical results can be applied to the real non-linear universe. We thus proceed to compare the density and velocity field skewness with observational data (IRAS and POTENT). The results are compatible with gaussian initial conditions, and local biasing.

## 1 The PDF as a Measure of Large Scale Structures

Many approaches can be used to characterize statistically the observed Large Scale Structures in the Universe. Here we focus on the one-point distribution function, or PDF. Empirically, this approach was first applied by Hubble (1934). Let $Q(r)$ be a physical quantity whose value depends on the location. Let $Q_\ell(r)$ be the smoothed value of $Q$ on scale $\ell$, *i.e.*, $Q_\ell(r)$ is obtained by convolving the scalar field $Q(r)$ by a smoothing window $W_\ell$ of characteristic scale $\ell$ ($Q_\ell(\mathbf{r}) = Q(\mathbf{s}) * W_\ell(\mathbf{r} - \mathbf{s}) =$

$\int d^3s\, Q(s)\, W_\ell(r-s))$. The probability that $Q_\ell$ lies between $Q$ and $Q+dQ$ is $P(Q,\ell)\, dQ$, $P$ being its PDF.

If $Q \equiv N$ is a number of galaxies, and $W$ is a top-hat window, then the PDF gives the probability of finding $N$ galaxies in a sphere (in 3D) of size $\ell$. In the following, $Q$ will be either the mass density contrast field $\delta = \rho/\overline{\rho} - 1$, or the divergence of the associated velocity field in units of the Hubble constant, $\theta = \nabla \cdot \mathbf{v}/H$; we will consider the two most commonly used types of windows, a top-hat and a gaussian one. Later on, we shall discuss the relation of the mass distribution to the distribution of luminous galaxies.

It is frequently assumed that, very early in the history of the Universe, the density contrast $\varepsilon \equiv \delta(t = t_i)$ could be taken as normally distributed, i.e.,

$$P(\varepsilon, \ell) \propto \exp\left[-\frac{\varepsilon_\ell^2}{2\,\sigma_i^2(\ell)}\right],$$

where $\sigma_i^2(\ell) \equiv \langle \varepsilon_\ell^2 \rangle$ stands for the initial variance of the density contrast field smoothed on scale $\ell$. It is assumed to be small at all scales, $\sigma_i \ll 1$. The second moment is the only quantity we need to know in order to fully characterize such a field. Indeed all odd moments of a gaussian PDF are zero, while the $2n$-th moments are proportional to the $n$ power of the variance, $\langle \varepsilon^{2n} \rangle \propto \sigma_i^{2n}$ (the corresponding specification in Fourier space is given in (7)). In the following, we shall refer to such a state as gaussian initial conditions.

The $P(\delta, \ell)$ moments are simply related to other statistical quantities. For instance,

$$\begin{aligned} \langle \delta_\ell^m \rangle &= \langle \delta(\mathbf{r}_1) * W_\ell(\mathbf{r} - \mathbf{r}_1)\, \delta(\mathbf{r}_2) * W_\ell(\mathbf{r} - \mathbf{r}_2) \ldots \delta(\mathbf{r}_m) * W_\ell(\mathbf{r} - \mathbf{r}_m) \rangle \\ &= \int d^3 r_1\, d^3 r_2 \ldots d^3 r_m\, W_\ell(\mathbf{r}_1) W_\ell(\mathbf{r}_2) \ldots W_\ell(\mathbf{r}_m)\, F_m(\mathbf{r}_1, \mathbf{r}_2 \ldots \mathbf{r}_m), \end{aligned}$$

that is the $m$-th order moment is the volume average (with weights $W_\ell$) of the full $m$-point correlation function, $F_m(\mathbf{r}_1, \mathbf{r}_2 \ldots \mathbf{r}_m) \equiv \langle \delta(\mathbf{r}_1) \delta(\mathbf{r}_2) \ldots \delta(\mathbf{r}_m) \rangle$ (since the overall homogeneity and isotropy of the system implies $F_m = \langle \delta(\mathbf{r}_1 - \mathbf{r}) \delta(\mathbf{r}_2 - \mathbf{r}) \ldots \delta(\mathbf{r} - \mathbf{r}_m) \rangle$). These correlation functions have been the most widely used indicators so far to describe the small-scale galaxy distribution; they give the probability of finding $m$ galaxies in infinitesimal volumes around $\mathbf{r}_1, \mathbf{r}_2 \ldots \mathbf{r}_m$; only the lowest orders ($m \leq 4$) are well-known (and the reduced parts of $F_2$, $F_3$, and $F_4$ are usually denoted (respectively) by $\xi \equiv F_2$, $\zeta \equiv F_3$, and $\eta(\mathbf{x}_1, \mathbf{x}_2, \mathbf{x}_3, \mathbf{x}_4) \equiv F_4 - F_2(\mathbf{x}_1, \mathbf{x}_2) F_2(\mathbf{x}_3, \mathbf{x}_4) - F_2(\mathbf{x}_1, \mathbf{x}_3) F_2(\mathbf{x}_2, \mathbf{x}_4) - F_2(\mathbf{x}_1, \mathbf{x}_4) F_2(\mathbf{x}_2, \mathbf{x}_3)$; see LSS for further details). Also, if the $m$-th order moment behaves as a power law of scale $\ell$, it can be used to obtain the Renyi indices $D_m$ of multifractal analysis ($\langle \delta_\ell^m \rangle \propto \ell^{(m-1)D_m}$, see e.g., [2] for further details). The (one-point) PDF gives only a partial statistical description of a system. A complete description calls for an infinite hierarchy of $m$-point PDFs (the $m$-point PDF is the probability of finding $\delta_\ell(1)$, $\delta_\ell(2)$, ... $\delta_\ell(m)$ at points $\mathbf{r}_1$, $\mathbf{r}_2$, ..., $\mathbf{r}_m$). Still we shall see that this simple object contains a wealth of useful information.

Our first task will be to derive the properties of the PDF today, under the assumption of a time-evolution under the sole influence of the gravitational instability acting on gaussian initial conditions. Under the influence of gravity, underdense regions become even more underdense (although not indefinitely, since the density is bounded from below), while positive density enhancements tend to grow without bound. Clearly the symmetry of the distribution cannot be maintained, and the PDF becomes skewed, i.e., $\langle \delta^3 \rangle$ departs from zero. The distribution also develops a non-zero kurtosis $\langle \delta^4 \rangle - 3\langle \delta^2 \rangle^2$. In the limit of a small variance, $\langle \delta^2 \rangle \ll 1$, the development of skewness and kurtosis are the most important effects.

In the following, we first recall how one obtains the lowest order moments of the PDF in the weakly non-linear regime in Eulerian perturbation theory. We then address the question of the $\Omega$ dependence of the results, and of smoothing. We show next the relationship between the low order moments and the overall shape of the PDF, in the case of gaussian initial conditions. We proceed by discussing the overall validity range of this approach. We also show that Lagrangian perturbation theory is well suited to compute the effect on the density field of using redshifts as distance estimates. We conclude by a discussion of the implications of these results by comparing with observational data.

## 2 Perturbation Theory

Here we follow the standard Eulerian approach, as described in Peebles' book [36, hereafter LSS] in §18. It is assumed that matter may be described as a non-relativistic pressureless fluid embedded in a Friedman-Lemaître model with zero cosmological constant.

### 2.1 Method

One starts from the standard equations for a perfect fluid, *i.e.*, the continuity equation, Euler equation, and Poisson equation. We write the perturbative expansion for the density contrast field as

$$\delta(\mathbf{x}, t) = \delta^{(1)}(\mathbf{x}, t) + \delta^{(2)}(\mathbf{x}, t) + \ldots,$$

where $\mathbf{x}$ are the comoving coordinates, and $t$ is the cosmological time. A similar expansion is performed on the proper peculiar velocity $\mathbf{v}$. The term $\delta^{(1)}$ is the linear order solution of the hydrodynamic equations of motion in comoving coordinates,

$$\delta^{(1)} = D(t)\varepsilon(\mathbf{x}) + \text{decaying mode},\tag{1}$$

where $\varepsilon$ depends on the spatial coordinates alone and $D(t)$ is the standard growing mode (LSS, eq. [11.16]). The term $\delta^{(2)} = O(\delta^{(1)})^2 = O(\varepsilon^2)$ is the solution of the equations of motion with quadratic nonlinearities included iteratively by using $\delta^{(1)}$ as source terms (as in LSS, §18). The dominant mode in the second order solution is

$$\delta^{(2)} = D(t)^2 \left[\frac{2}{3}(1+\kappa)\varepsilon^2 + \nabla\varepsilon \cdot \nabla\phi + (\frac{1}{2} - \kappa)\tau^2\right],\tag{2}$$

where $\nabla = \partial/\partial\mathbf{x}$ while $\phi(\mathbf{x}) = -\int d^3x' \varepsilon(\mathbf{x}')/4\pi|\mathbf{x} - \mathbf{x}'|$ is the linearized Newtonian gravitational potential, and

$$\tau^2 = \sum_{\alpha,\beta=1}^{3} (\tau_{\alpha\beta})^2, \qquad \text{with} \quad \tau_{\alpha\beta} = (\tfrac{1}{3}\delta_{\alpha\beta}\nabla^2 - \nabla_\alpha\nabla_\beta)\phi.$$

Here $\alpha, \beta$ are the indices of the spatial coordinate components, $\mathbf{x} = \{x_\alpha\}$. Apart from a multiplicative factor, $\tau_{\alpha\beta}$ is the linear order peculiar velocity shear tensor (LSS, eq. [14.12]). The parameter $\kappa(t)$ is a slowly varying function of cosmological time. Bouchet *et al.* (1993, Ref. [8]) show that it is well approximated by

$$\kappa \approx \tfrac{3}{14}\Omega^{-2/63},\tag{3}$$

in the range $0.05 \geq \Omega \geq 3$ (the accuracy of this approximation is then better than 0.4%). For $\Omega = 0$, $\kappa = \tfrac{1}{4}$. The exact expression for $\kappa(\Omega)$, valid in the entire range $\Omega \geq 0$, is given in [8]). For $\Omega = 1$, $\kappa = \tfrac{3}{14}$, and we recover the well known Einstein-de Sitter solution (*e.g.*, LSS, eq. [18.8] or Ref. [25]).

### 2.2 Skewness for the Density Field

Now we can calculate the gravitationally induced skewness under the assumption that $\varepsilon$ is a random gaussian field. The lowest order terms in the series for $\langle\delta^3\rangle$ are

$$\langle\delta^3\rangle = \langle\delta^{(1)\,3}\rangle + \langle 3\,\delta^{(1)\,2}\delta^{(2)}\rangle + O(\varepsilon^5).\tag{4}$$

The linear solution (1) implies that the first term is $D(t)^3$ times the initial skewness, which is zero for gaussian initial conditions. The second order solution (2) shows that the second term involves $\langle\varepsilon^4\rangle$, which is $\propto \sigma^4$ for gaussian initial conditions. Thus the skewness ratio

$$S_3 \equiv \frac{\langle\delta^3\rangle}{\langle\delta^2\rangle^2}\tag{5}$$

is a constant versus scale, and one finds (ref. [8]) up to terms of order $\varepsilon^2$ (since $\langle\varepsilon^5\rangle = 0$),

$$S_3 = 4 + 4\kappa(\Omega) \simeq \frac{34}{7} + \frac{6}{7}(\Omega^{-2/63} - 1) \ . \tag{6}$$

The $\simeq$ sign above applies to the range of applicability of equation (3). The first term of this equation, 34/7, had been obtained by Peebles more than a decade ago (LSS, §18). The weak $\Omega$-dependence of the full expression shows that nearly all the $\Omega$-dependence of the skewness $\langle\delta^3\rangle$ comes from that of the square of the variance. It simply reflects the fact that the second order growth rate is nearly equal to $D(t)^2$, as can be seen from (2).

### 2.3 Smoothing

To make contact with observables, we want the skewness of the density field $\delta_\ell$, when *smoothed* with either a top-hat or a gaussian (spherically symmetric) window, which satisfies

$$\int W(x)\,d^3x = 1, \quad \text{and} \quad \int W(x)\,x^2\,d^3x = \ell^2 \ .$$

The first equation insures a proper normalization to unity, while the second requires the effective half-width $\ell$ to be finite. The top hat case is appropriate for comparisons with the observed frequency distribution of galaxies. (In that case, discreteness corrections must be taken into account before comparing with the theory, *e.g.*, $\langle(N/\overline{N} - 1)^2\rangle = 1/\overline{N} + \langle\delta^2\rangle$ if one adopts the Poisson model, see *e.g.*, LSS §33. A gaussian window, on the other hand, removes these discreteness fluctuations, since it does not have sharp edges.) Here we recall the main results of [27].

The calculations are most conveniently done in Fourier space. For the initial field $\varepsilon$, the Fourier components are given by

$$\varepsilon_\mathbf{k} \equiv \frac{1}{(2\pi)^{3/2}} \int \varepsilon(\mathbf{x}) \exp(i\mathbf{k}\cdot\mathbf{x})\,d^3x \ .$$

For gaussian initial conditions, one has

$$\langle\varepsilon_\mathbf{k}\varepsilon_{\mathbf{k}'}\rangle = \delta_D(\mathbf{k}+\mathbf{k}')P(k), \qquad \langle\varepsilon_\mathbf{k}\varepsilon_{\mathbf{k}'}\varepsilon_\mathbf{q}\rangle = 0, \tag{7}$$

$$\langle\varepsilon_\mathbf{k}\varepsilon_{\mathbf{k}'}\varepsilon_\mathbf{q}\varepsilon_{\mathbf{q}'}\rangle = P(k)P(q)\delta_D(\mathbf{q}+\mathbf{q}')\delta_D(\mathbf{k}+\mathbf{k}') + \text{cycl. (two terms)} \ ,$$

where $\delta_D$ is the Dirac delta, and $P(k)$ is the (initial) power spectrum. Now we can use the above expressions for the first few moments, together with equations (1), (2), and (4) to derive $S_3$ to lowest non-vanishing order,

$$S_3 = \int \frac{d^3k\,d^3k'}{(2\pi)^6\,\sigma^4}\,P(k)\,P(k')\,W_k\,W_{k'}\,W_{|\mathbf{k}-\mathbf{k}'|}\,T(\mathbf{k},\mathbf{k}') + O(\sigma^2) \ . \tag{8}$$

Here $T(\mathbf{k},\mathbf{k}')$ stands for $T(\mathbf{k},\mathbf{k}') = 4 + 4\kappa(\Omega) - 6\mu(k/k') + [2 - 4\kappa(\Omega)]\,P_2(\mu)$, where $\mu = \mathbf{k}\cdot\mathbf{k}'/kk'$, and $P_2$ is a Legendre polynomial. If there is no smoothing ($W_\mathbf{k} = 1$), the dipole and quadrupole terms integrate to zero, and one simply recovers (6). On the other hand, as soon as one introduces smoothing, the result does depend on the initial power spectrum $P(k)$.

Let us assume that $P(k) \propto k^n$. Then, for a top-hat smoothing, and $-3 \leq n < 1$, the equation (8) yields after painful calculations the simple result

$$S_3 = 4 + 4\kappa(\Omega) - (3 + n) \ , \tag{9}$$

A similar feat can also be accomplished for a gaussian window, although the result is slightly less simple (see figure 4). Actually, a careful inspection of the expression (8) shows that $S_3$ should only depends on the effective (logarithmic) slope of the power spectrum at the smoothing scale. This was confirmed in [27] by comparing numerical integration for a CDM power spectrum with a prediction using (9). Our result was recently generalized by Bernardeau in [5], who showed that for an arbitrary gaussian field, smoothed with a top hat filter on a scale $\ell$,

$$S_3(\ell) = 4 + 4\kappa(\Omega) - \gamma_\ell, \quad \text{with} \quad \gamma_\ell = -\frac{\partial \log \langle \delta^2 \rangle}{\partial \log \ell}. \tag{10}$$

For a pure power law $P(k)$, we have $\gamma = 3 + n$, in agreement with (9).

## 2.4 Skewness for the Velocity Field

The continuity equation gives the divergence of the velocity field in terms of the time derivative of the density contrast. Let us call $T_3$ the skewness ratio for the divergence of the velocity field in units of the Hubble constant ($\theta = \nabla \cdot \mathbf{v}/H$). One then finds (ref. [6]; see also equation (28) in §2.7.2), for a top-hat smoothing,

$$T_3 \equiv \frac{\langle \theta^3 \rangle}{\langle \theta^2 \rangle^2} = -\Omega^{-0.6} \left[ \frac{26}{7} - \gamma_\ell \right]. \tag{11}$$

The asymmetry in the distribution of $\theta$ is directly related to the asymmetry in $\delta$: voids and clusters in the mass distribution correspond to sources ($\theta > 0$) and sinks ($\theta < 0$) in the velocity field.

## 2.5 Shape of the PDFs and Edgeworth Expansion

We now wish to examine how gravitational instability drives a PDF away from its initial state, which we assume to be gaussian. We start by the density field PDF, $P(\delta, \ell)$. We need to introduce the Gram-Charlier expansion, which allows one to reconstruct the PDF from its moments. Then we rearrange the Gram-Charlier series by collecting all terms of the same order. The result is the proper asymptotic expansion of the PDF in powers of $\sigma_\ell$, that we introduced in [28].

Let us consider $p(\nu)$, the PDF of the density field in terms of the standardized random variable $\nu \equiv \delta_\ell/\sigma_\ell$. Let us also introduce $\phi(\nu) = (2\pi)^{-1/2} \exp(-\nu^2/2)$, a gaussian (or Normal) PDF. Since we want to describe an evolution from gaussian initial conditions, it makes sense to consider an expansion of $p(\nu)$ in terms of $\phi(\nu)$ and its derivatives. The *Gram-Charlier series* (Cramér 1946 and references therein) provides such an expansion:

$$p(\nu) = c_0 \phi(\nu) + \frac{c_1}{1!} \phi^{(1)}(\nu) + \frac{c_2}{2!} \phi^{(2)}(\nu) + \ldots, \tag{12}$$

where $c_m$ are constant coefficients. Superscripts denote derivatives with respect to $\nu$:

$$\phi^{(m)}(\nu) \equiv \frac{d^m \phi}{d\nu^m} = (-1)^m H_m(\nu) \phi(\nu), \tag{13}$$

where $H_m$ is the Hermite polynomial of degree $m$.

The Hermite polynomials satisfy orthogonality relations (e.g. Abramowitz & Stegun 1964):

$$\int_{-\infty}^{\infty} H_m(\nu) H_p(\nu) \phi(\nu) \, d\nu = \begin{cases} 0, & \text{if } m \neq p \, ; \\ m!, & \text{otherwise.} \end{cases}$$

Therefore, multiplying both sides of equation (12) by $H_m$ and integrating term by term yields

$$c_m = (-1)^m \int_{-\infty}^{\infty} H_m(\nu) \, p(\nu) \, d\nu. \tag{14}$$

Equation (14) gives $c_0 = 1$, $c_1 = c_2 = 0$, while for the next four coefficients in the series we obtain

$$c_m = (-1)^m S_m \sigma_\ell^{m-2}, \quad \text{for } 3 \leq m \leq 5 \, ; \quad c_6 = S_6 \sigma_\ell^4 + 10 S_3^2 \sigma_\ell^2. \tag{15}$$

Thus the $S_m(\ell)$ have both a dynamic and a static application: they describe the time evolution of moments of the PDF at a fixed smoothing scale $\ell$, and they also describe the relation between moments of the PDF at a fixed time on different smoothing scales. In the latter case, one must also include the scale-dependence of the $S_m$ if the initial power spectrum is not scale-free.

We have just seen that perturbation theory and gaussian initial conditions imply that $\langle \delta_\ell^3 \rangle \propto \sigma_\ell^4$, and $S_3$ is therefore an "order unity" quantity when $\sigma_\ell$ is the "small" parameter. The same is true for all remaining reduced moments, $S_m = O(1)$ for all $m$ (Bernardeau 1992, Fry 1994). For our immediate purpose here, the important consequence of this is that the Gram-Charlier series *is not* a proper asymptotic expansion for $p(\nu)$. In an asymptotic expansion, the remainder term should be of higher order than the last term retained. However, if we truncated the series (12) at the $\phi^{(4)}$ term, which is $O(\sigma^2)$, we would miss another $O(\sigma^2)$ contribution coming from $c_6$ (equation (15)). In order to deal with this problem, let us rearrange the Gram-Charlier expansion by collecting all terms with the same powers of $\sigma$. The result is the so-called *Edgeworth series*, with the first few terms given by

$$p(\nu) = \phi(\nu) - \frac{1}{3!} S_3 \phi^{(3)}(\nu) \sigma + \frac{1}{4!} S_4 \phi^{(4)}(\nu) \sigma^2 + \frac{10}{6!} S_3^2 \phi^{(6)}(\nu) \sigma^2 + O(\sigma^3) \ . \tag{16}$$

Cramér (1946) lists the Edgeworth series to higher order, and he proves that it is a proper asymptotic expansion. This proof is directly relevant to our purposes, since it implies that there are no additional $O(\sigma^2)$ terms hiding in the Gram-Charlier series at $m > 6$.

Now we can see the attractiveness of the Edgeworth series for describing the gravitational evolution of gaussian fluctuations: it becomes a series expansion for the evolving PDF in powers of the r.m.s. fluctuation $\sigma$. This makes physical sense because the Edgeworth series provides an expansion about a gaussian probability distribution. If the initial fluctuations are gaussian, then we expect the terms describing successively larger departures from a gaussian PDF to come in with successively higher powers of $\sigma$. For similar reasons, the Edgeworth expansion has recently found applications in stellar dynamics as a description of galaxy line profiles (e.g. van de Van der Marel & Franx 1993; Gerhard 1993). A multivariate Edgeworth expansion is also well-known in kinetic theory: this is the so-called Grad solution of the Boltzmann equation (Klimontovich 1982).

Given equation (16), we can compute the Edgeworth approximation to the PDF provided that we can compute $S_m$ to the required order. In this paper we will make use of the second-order approximation,

$$p(\nu) = \left[ 1 + \frac{1}{3!} S_3 \sigma H_3(\nu) \right] \phi(\nu), \tag{17}$$

and the third-order approximation,

$$p(\nu) = \left[ 1 + \frac{1}{3!} S_3 \sigma H_3(\nu) + \frac{1}{4!} S_4 \sigma^2 H_4(\nu) + \frac{10}{6!} S_3^2 \sigma^2 H_6(\nu) \right] \phi(\nu). \tag{18}$$

Although equation (17) contains only a single explicit power of $\sigma$, it is appropriately described as a second-order approximation because the parameter $S_3$ remains zero until second order in perturbation theory. Similarly, equation (18) is a third-order approximation because $S_4$ remains zero until third order.

Of course, if the variable we are interested in is $\theta$ instead of $\delta$, one just adopts the appropriate analogues of $S_m$, *i.e.*, $S_3 \longrightarrow T_3 \equiv \langle \theta^3 \rangle \langle \theta^2 \rangle^{-2}$, and so on. Similarly, if we are interested in the PDF shape in redshift space, one just needs to use the corresponding analogue of $S_m$, $S_m^z$, see §2.7.2). The Edgeworth series may also be used to relate $S_3$ to other measures of asymmetry like $\langle \delta | \delta | \rangle$ which, according to Nusser & Dekel (1993), may offer better signal to noise ratio than $\langle \delta^3 \rangle$ when applied to real galaxy surveys (by being less sensitive to the tail of the PDF, *i.e.*, to rare events). This latter quantity is not easy to compute directly by perturbation theory (see the appendix of [28]), but it is trivial to obtain

$$\langle \delta | \delta | \rangle = \sqrt{\frac{2}{9\pi}} S_3 \sigma^3 + O(\sigma^5) \ . \tag{19}$$

by using the approximation (18).

## 2.6 Validity of Perturbation Theory

The previous results assume that the system never gets to be strongly non-linear, *i.e.*, $\sigma_\ell \ll 1$ is true at *every* scale. But in practice we know that the observed density contrast is very large at small scales. One only gets $\sigma_\ell \ll 1$ for $\ell \gg 8h^{-1}$ Mpc. It is thus by no means obvious that there is any range today for which perturbation theory might be applicable. On the other hand, it has long ago been noticed that linear perturbation theory yields a good description of the variance [or the 2-body correlation function $\xi(t) \propto D(t)^2/D(t_i)\,\xi(t_i)$] measured in fully non-linear numerical simulation. Actually, the agreement remains true even for values of the variance as large as $\sim 2$!

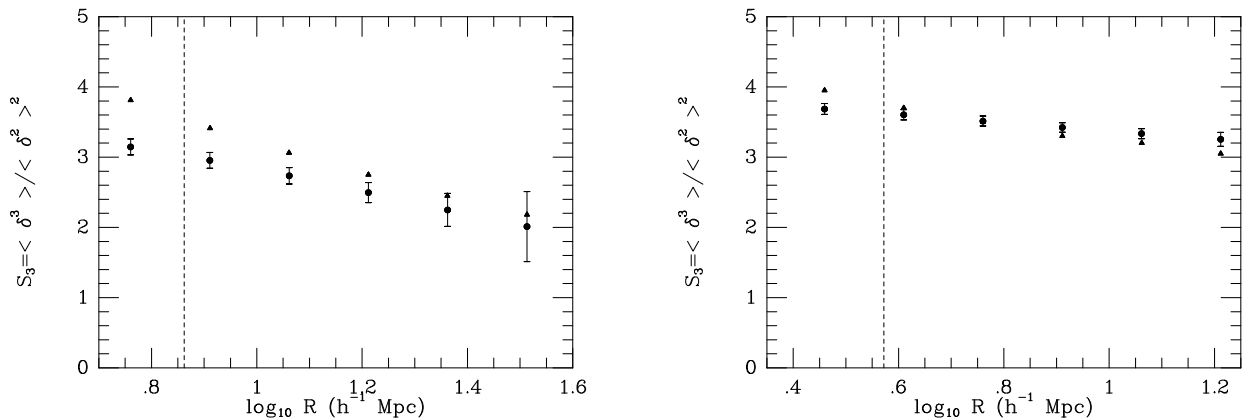

Figure 1: The measured $S_3(\ell)$ in a CDM simulation (triangles) is compared to the theory (circles; the error bars come form the numerical uncertainty in evaluating the integrals giving $S_3$). The left panel corresponds to a spherical top-hat smoothing, while the right one corresponds to a gaussian smoothing. The vertical dashes mark the limit between the strongly non-linear regime ($\sigma \geq 1$) and the weakly non-linear one ($\sigma \leq 1$. Courtesy Colombi 1993 & Juszkiewicz *et al.* 1993a (Ref. [12] & [27]).

It turns out that what holds true for linear perturbation theory results is also true at higher orders. Indeed, our top-hat results for $S_3$ have been checked against the simulation results of Efstathiou *et al.* 1988, Bouchet and Hernquist 1992, and Lahav *et al.* 1993 (Refs. [15], [9], & [31]) in the case of scale-free initial spectra. In the case of a gaussian smoothing, we checked the theory with the scale free-simulations of Weinberg and Cole 1991. In the CDM case, comparison were made with the results in Bouchet and Hernquist 1992 (top-hat case, Ref. [9]), as well as those we obtained for that purpose which are displayed in figure 1. In all those cases, the agreement between the perturbation theory and N-body experiments was excellent (see [27] for further details).

An analysis of $P(\delta,\ell)$ and $P(\theta,\ell)$ can be found in Juszkiewicz *et al.* 1993b (Ref. [28] from which we extracted figure 2). It shows the results for a scale-free, $n = -1$, case analyzed with a gaussian smoothing. On the left is shown the evolution of the asymmetry measure $M_3 \equiv \langle \delta^3 \rangle$ and $\widetilde{M_3} \equiv \langle \delta | \delta | \rangle$. Logarithms are base-10. In the lower panel, the solid line shows the prediction of second-order perturbation theory, $M_3 = S_3 \sigma^4$, using the value $S_3 = 3.47$ (appropriate to an $n = -1$ power spectrum and gaussian smoothing filter). Points show measurements from the density fields of the N-body simulations, with smoothing lengths of 2, 4, and 8 cells (circles, triangles, and squares, respectively). For closer inspection, the upper panel plots the ratios $M_3/\sigma^4$ (top points) and $\widetilde{M_3}/\sigma^3$ (bottom points), with horizontal lines representing the analytic predictions. Error bars mark the $1\sigma$ theoretical uncertainty, *i.e.*, the run-to-run dispersion of eight independent simulations divided by $7^{1/2}$. The right side of the figure shows the evolution of $-M_{3\theta} \equiv -\langle \theta^3 \rangle$ and $-\widetilde{M_{3\theta}} \equiv -\langle \theta | \theta | \rangle$. To plot the lines, we use the appropriate value $T_3 = -2.19$, so there are no free parameters to either of these "fits". It demonstrates (see also figure 7) that perturbative and N-body results agree to within the $1\sigma$ uncertainty when $\sigma$ is small, as expected.

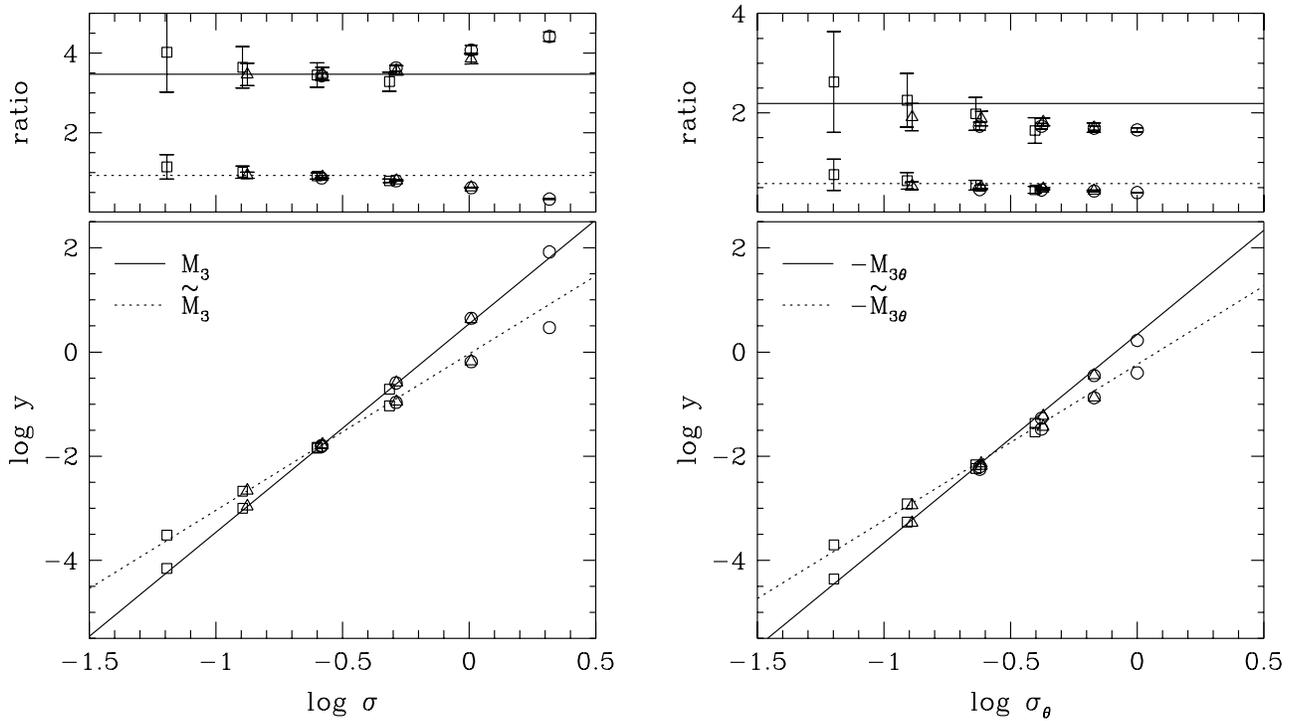

Figure 2: See legend in text. Courtesy Juszkiewicz *et al.* 1993b (Ref. [28])

Note also that the N-body results for $M_3$ remain remarkably close to the perturbation theory prediction even when $\sigma = 1$. A similar applicability range was found in the other studies mentioned so far. Although these numerical comparisons are not general analytical proofs, they do show the usefulness of perturbation theory for the cases of interest in the cosmological context.

One can also check when the Edgeworth series and perturbation theory are appropriate to describe the overall shape of the PDF, by comparing its predictions with the results of N-body experiments. The solid lines of figure 3 show the PDF of the smoothed density fields of matter density (left column) and velocity divergence (right column), at three different stages in the simulations used above (to produce figure 2). The corresponding *rms* amplitude ($\sigma$ and $\sigma_\theta$) are shown in each panel, and the variables on the horizontal axes are $\nu = \delta/\sigma$ and $\nu = \theta/\sigma_\theta$. In all the panels, dotted and dashed lines show the approximations corresponding to equations (17) and (18) respectively. The approximations work well for $S_3\sigma < 1$ and $|\nu\sigma| < 1$ (and equivalent requirements for $\theta$). They begin to break down outside of that range, as expected.

## 2.7 Lagrangian versus Eulerian Approach

As was shown above, a lot can be accomplished by using the standard Eulerian perturbative approach to gravitational instability. However, perturbative equations of motion are often easier to integrate when expressed in Lagrangian coordinates. In our experience, this happened at second order for $\Omega \neq 1$ (see §2.7.1 below). Another example is the redshift space distortion for $\langle \delta^3 \rangle$, discussed in §2.7.2. Moreover, as $\langle \delta^2 \rangle$ grows with time, at any fixed order in perturbation theory, the Lagrangian approach is likely to remain valid longer than the Eulerian approach. This is so because the requirement of small Lagrangian displacements and gradients is weaker than the requirement $\langle \delta^2 \rangle \ll 1$. This idea, which motivated the Zel'dovich (1970) approximation (which is simply the first order Lagrangian solution), remains valid at higher orders.

In the following, we just outline the derivation of the perturbative solutions, and focus on the redshift distortion problem. We also give examples of comparison between Eulerian and Lagrangian theory when $\langle \delta^2 \rangle \simeq 1$.

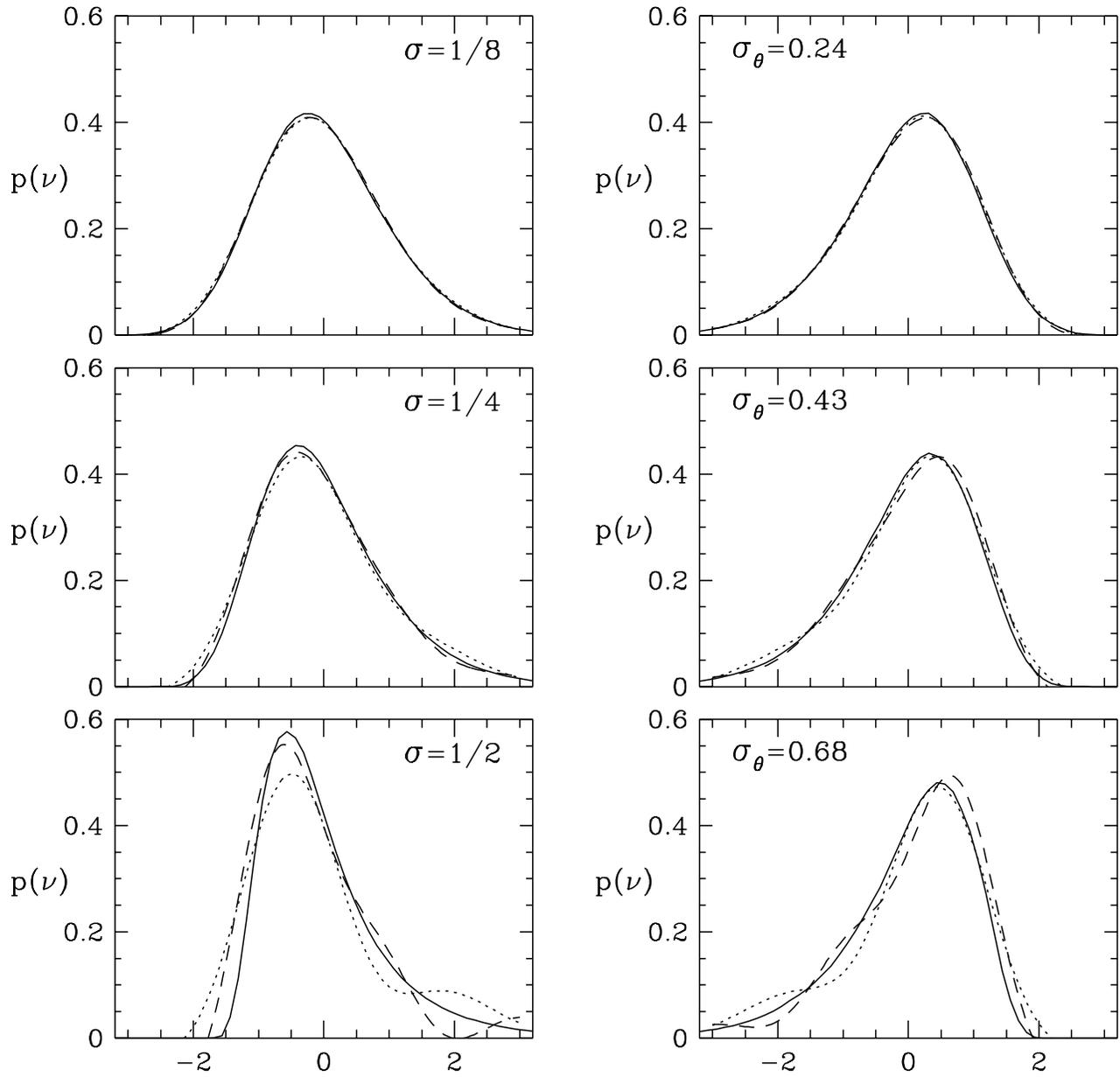

Figure 3: See legend in text. Courtesy Juszkiewicz *et al.* 1993b (Ref. [28])

*2.7.1 Lagrangian Formalism* In the Lagrangian approach, the primary object of the analysis is the particle trajectory instead of the density contrast. A fluid element (a "particle") is indexed by its unperturbed Lagrangian coordinate **q** and its comoving Eulerian position at time t, $\mathbf{x}(t)$, is connected to **q** by a displacement field $\Psi$

$$\mathbf{x} = \mathbf{q} + \Psi(t, \mathbf{q}). \tag{20}$$

An equation for $K = \nabla \cdot \Psi$ may be obtained by taking the divergence of the motion equation, which involves (once Poisson equation is used) the density contrast. The latter is expressed by requiring that mass be conserved, *i.e.*,

$$\rho(\mathbf{x}) J \, d^3q = \rho(\mathbf{q}) d^3q.$$

Thus $\delta = J^{-1} - 1$, if $J$ stands for the determinant of the Jacobian of the transformation from **x** and **q**. Of course $\Psi$ is then only determined up to the addition of any divergence-free displacement field[1].

As in the Eulerian case, perturbative solutions are obtained by means of an iterative procedure, as seems to have been done for the first time by Moutarde *et al.* (1991; it was further developed by Bouchet *et al.* 1992; 1993, refs.[8] & [11]; it also initiated a parallel effort by Buchert and collaborators, see the contribution of Weiss and Buchert in this volume, and references therein). Let us denote by $K^{(m)}$, $L^{(m)}$, and $M^{(m)}$, the $m - th$ order part of the (invariant) scalars

$$\begin{cases} K &= \nabla \cdot \Psi = \sum_l \Psi_{l,l} \\ L &= \frac{1}{2} \sum_{l \neq m} (\Psi_{l,l} \Psi_{m,m} - \Psi_{l,m} \Psi_{m,l}) \\ M &= |det[\Psi_{i,j}]| \end{cases} \tag{21}$$

where $\Psi_{l,m}$ denotes $\delta \Psi_l / \delta q_m$, *i.e.*, the partial derivative of the $l - th$ component of the displacement field with respect to the $m - th$ component of the lagrangian coordinate. Focusing on the fastest growing mode solutions, we then find

$$\begin{aligned} K(t, \mathbf{q}) &= g_1(t) \, K^{(1)}(t_i, \mathbf{q}) + g_2(t) \, L^{(2)}(t_i, \mathbf{q}) \\ &+ g_{3a}(t) \, M^{(3)}(t_i, \mathbf{q}) + g_{3b}(t) \, L^{(3)}(t_i, \mathbf{q}) \end{aligned} \tag{22}$$

where $g_1(t)$ is nothing else than the standard linear Eulerian solution $D(t)/D(t_i)$, and $g_2$, $g_{3a}$, and $g_{3b}$ behave near $\Omega = 1$ as

$$g_2 \approx -\frac{3}{7} \Omega^{-2/63} g_1^2, \quad g_{3a} \approx -\frac{1}{3} \Omega^{-4/77} g_1^3, \quad g_{3b} \approx -\frac{10}{21} \Omega^{-2/35} g_1^3.$$

Further details, and expressions for flat models with $\Lambda \neq 0$ are given in [11].

The first (linear) term of (22) is nothing else than Zel'dovich solution, since equation (22) implies for a potential movement,

$$\Psi^{(1)}(\tau, \mathbf{q}) = g_1(t) \, \tilde{\Psi}^{(i)}(\mathbf{q}), \tag{23}$$

if we denote by $\tilde{\Psi}^{(i)}(\mathbf{q})$ the initial displacement field. Also, note that $\delta^{(1)}(t) = -\nabla \cdot \Psi^{(1)} \propto g_1(t)$, *i.e.*, the Eulerian linear behavior is recovered. The second order growth rate, $g_2$, yields the expression for $\kappa(\Omega)$ introduced earlier [eq. (3)]. And at this order, the solution is again separable as a growth rate times a purely spatial part (denoted by tilde) which is set at the initial time:

$$\Psi^{(2)} = g_2(t)\tilde{\Psi}^{(2)}(\mathbf{q}), \quad \text{with} \quad \nabla \cdot \tilde{\Psi}^{(2)}(\mathbf{q}) = L^{(2)}(t_i) = \frac{1}{2} \sum_{l \neq m} (\tilde{\Psi}^{(i)}_{l,l} \tilde{\Psi}^{(i)}_{m,m} - \tilde{\Psi}^{(i)}_{l,m} \tilde{\Psi}^{(i)}_{m,l}) \tag{24}$$

(for a potential movement).

---
[1] When needed, one may lift this indeterminacy by restricting one's attention to potential movements, which must satisfy $\nabla_\mathbf{x} \times \ddot{\mathbf{x}} = 0$. This case is relevant, since vortical perturbations decay with time in linear perturbation theory, a consequence of the conservation of angular momentum in an expanding universe. Thus one might consider that such solutions will apply, if vorticity is initially present, at later times when it has decayed away.

*2.7.2 Real Space-Redshift Space Mapping* In redshift space, the appearance of structures is distorted by peculiar velocities. At "small" scales, this leads to the "finger of god" effect: the clusters are elongated along the line-of-sight due to their internal velocity dispersion. This is an intrinsically non-linear effect, and we shall not be concerned with it. At "large" scales, the effect is reversed: the coherent inflow leads to a density contrast increase parallel to the line-of sight. Indeed, foreground galaxies appear further than they are, while those in the back look closer, both being apparently closer to the accreting structure (Sargent & Turner 1977; LSS, §76; Kaiser 1987).

Let us now calculate $\langle \delta^2 \rangle$ and $\langle \delta^3 \rangle$ using the Lagrangian approach. Since the unsmoothed density contrast is given by $\delta = J^{-1} - 1$, the first terms of its expansion will be given by $\delta^{(1)} = -J^{(1)}$, $\delta^{(2)} = J^{(1)\,2} - J^{(2)}$. We assume an initially gaussian density field. Thus we shall require that the three components of the initial displacement field $\tilde{\Psi}^{(i)}$ be independent and gaussian, which will insure that the density contrast is also gaussian since it is related by a linear operator to the displacement field ($\delta_i = -\nabla \cdot \tilde{\Psi}^{(1)}$). In that case, variance and skewness are given by

$$\langle \delta^2 \rangle = \langle J^{(1)\,2} \rangle + \mathcal{O}(\varepsilon^4), \quad \langle \delta^3 \rangle = \langle 2 J^{(1)\,4} - 3 J^{(1)\,2} J^{(2)} \rangle + \mathcal{O}(\varepsilon^6),$$

where all averages on the displacement field are taken with respect to the Lagrangian unperturbed coordinate[2] $\mathbf{q}$. The previous formulae take into account the fact that $\langle J^{(1)} \rangle = \langle J^{(2)} \rangle = \langle J^{(3)} \rangle = 0$, which also insures that $\langle \delta \rangle = 0$), as well as $\langle J^{(1)\,3} \rangle = 0$ (indeed $\langle \tilde{\Psi}^{(1)\,2m+1} \rangle = 0$).

If we call $\sigma_1^2$ the variance of any component $i$ of the gradient field $\sigma_1^2 = \langle \Psi_{i,i}^{(1)\,2} \rangle^2 \; (= g_1^2 \langle \varepsilon^2 \rangle^2 /3)$, which is also gaussian, we have $\langle J^{(1)\,4} \rangle = 27\sigma_1^4 = 3 \langle J^{(1)\,2} \rangle^2$. The other term in $\langle \delta^3 \rangle$ involves the product $J^{(1)\,2} J^{(2)}$ which can readily be estimated since, after the second order solution, we have $J^{(2)} = (1 + g_2/g_1^2) L^{(2)}$. It follows by development that $\langle J^{(1)\,2} J^{(2)} \rangle = 6(1 + g_2/g_1^2)\sigma_1^4$. We thus have the remarkably simple result

$$S_3 = 4 - 2 g_2/g_1^2 + \mathcal{O}(\varepsilon^2), \tag{25}$$

which is identical to (6). The first term corresponds to the pure Zel'dovich approximation and had been found by Grinstein and Wise (1987).

Let us now consider the case of spherical coordinates, when distances to the observer would be estimated by means of redshift measurements. And let us now denote redshift space measurements by the superscript $z$. The redshift space comoving position $\mathbf{x}^z$ of a particle located in $\mathbf{r}(\mathbf{q}) = a\mathbf{x}(\mathbf{q})$ is $\mathbf{x}^z = \dot{\mathbf{r}}/(aH)$ (with $H = \dot{a}/a$, where $a$ is the scale factor, while the dot represents a time derivative). The real space perturbative expansion (20) is then replaced by

$$\mathbf{x}^z = \mathbf{q} + [1 + f_1(t)] g_1(t) \tilde{\Psi}^{(1)}(\mathbf{q}) + [1 + f_2(t)] g_2(t) \tilde{\Psi}^{(2)}(\mathbf{q}) + \mathcal{O}(\varepsilon^3),$$

where we have explicitly used the separability of $\Psi^{(1)} = g_1(t) \tilde{\Psi}^{(1)}(\mathbf{q})$ and $\Psi^{(2)} = g_2(t) \tilde{\Psi}^{(2)}(\mathbf{q})$ (eqs. (23) & (24)). For $\Omega$ close to 1, the logarithmic derivatives of the growth rates are well approximated by

$$f_1 \equiv (a/g_1) \partial g_1/\partial a \approx \Omega^{3/5}, \quad \text{and} \quad f_2 \equiv (a/g_2) \partial g_2/\partial a \approx 2\, \Omega^{4/7}, \tag{26}$$

with $f_1$ taken from Peebles (1976), and the second from [11]. In the limit of an infinitely remote observer, say along the $r_3$-axis, the observed density constrast $\delta_z$ in comoving coordinates is simply $\delta^z(x_1, x_2, x_3^z)$, which amounts to approximate spherical coordinates by cartesian ones. All we have to do, then, is to replace everywhere in the calculation of $S_3$ the quantity $\Psi_3^{(m)} = g_m \tilde{\Psi}_3^{(m)}$ by $(1 + f_m) g_m \tilde{\Psi}_3^{(m)}$ (for $m = 1$ and 2). It follows that[3]

$$S_3^z = 6 - 6 \frac{1 + 2(1 + f_1)^2 + (g_2/g_1^2)\,[3 + 2f_1 + f_2 + f_1 f_2 \mathcal{E}]}{[2 + (1 + f_1)^2]^2}, \quad \text{with } 1 \geq \mathcal{E} \geq 0. \tag{27}$$

---

[2] Indeed, $\langle Q[x] \rangle_x = \langle Q[x(q)] J \rangle_q$ where the subscript $x$ and $q$ refer (respectively) to averages in Eulerian or Lagrangian space. This follows if one assumes, as usual, that ensemble averages $\langle ... \rangle$ are equivalent to averages over space, provided the volume is large enough that it can be considered a "fair" sample.

[3] All-symmetry breaking terms, which cannot be evaluated without specifying the initial power spectrum, are gathered in $\mathcal{E}$. This result was obtained by using Fourier analysis.

Further details of the derivation may be found in [11]. Of course, we recover the real space result (25) if we set $f_1 = f_2 = 0$. On the other hand, if $\Omega = 1$, we have $f_1 = 1 = f_2/2$ (and $g_2/g_1^2 = -3/7$), which yields $S_3 = (35 + \mathcal{E})/7$ while, for $\Omega = 0.1$, $S_3 \approx (34.5 + 0.4\mathcal{E})/7$. Since we used an "infinitely remote observer" approximation, the formula (27) strictly applies only in the limit of large volumes. In any case, at least in that limit, it clearly shows that the ratio $S_3$ is nearly independent of the value of $\Omega$, nor is it affected by redshift space distortions.

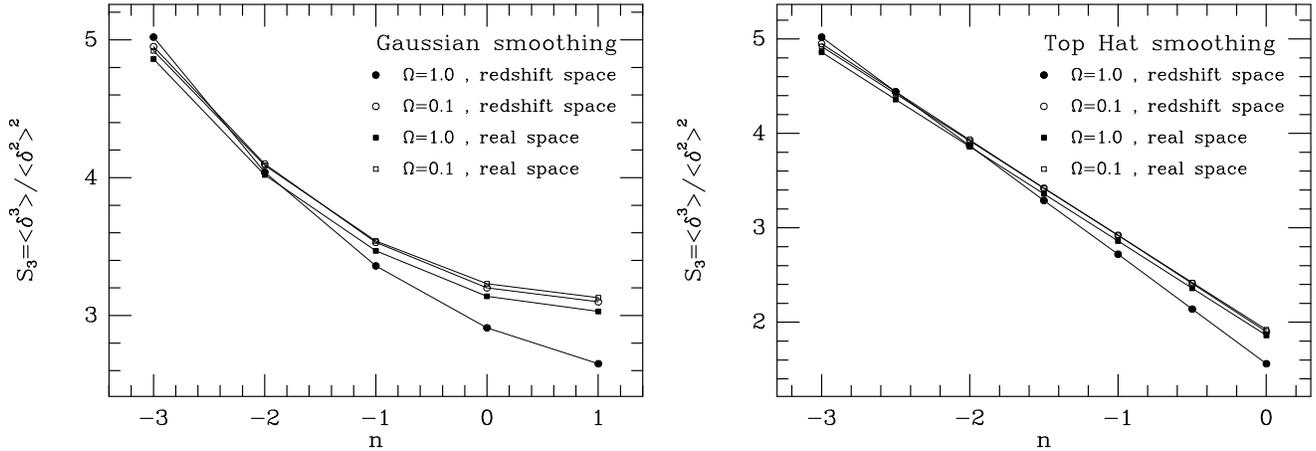

Figure 4: See legend in text. Courtesy Hivon *et al.* 1993 (Ref. [23])

We have recently extended those calculations by including smoothing and by dropping the simplifying assumption of an infinitely remote observer (*i.e.*, without approximating spherical coordinates by cartesian ones, but keeping the "large sample limit", as in [29]). The results are displayed in figure 4, which is extracted from [23]. They confirm that the skewness of an initially gaussian field is essentially the variance squared times a coefficient that depends mostly on smoothing and the initial variance. Most of the redshift or $\Omega$ dependence is contained in the variance dependence.

Finally, let us mention that $T_3$ can be obtained by similar techniques. Indeed, with the same notations than above

$$\nabla_q \cdot \left(\frac{\dot{\mathbf{x}}}{H}\right) = f_1 g_1 K^{(1)} + f_2 g_2 K^{(2)}.$$

Since $\theta^{(1)} = f_1 g_1 K^{(1)}$, and $\theta^{(2)} = f_2 g_2 K^{(2)} - f_1(\nabla_q \Psi^{(1)} \cdot \nabla_q \Psi^{(1)})$, the assumption of gaussian initial conditions in Fourier space (see note 2) yields

$$T_3 = -\frac{2}{f_1}\left(1 - \frac{f_2}{f_1}\frac{g_2}{g_1^2}\right) \simeq -\frac{26}{7}\Omega^{-0.6}. \qquad (28)$$

Once the expressions (26) are inserted in this formula, one recovers equation (11) when no smoothing correction is included[4]. By using Zel'dovich approximation, one would not include the second order term, which then yields $T_3^Z \simeq -2\Omega^{-0.6}$. When a top hat smoothing is included, Zel'dovich approximation yields $T_3^Z \simeq -\Omega^{-0.6}(n+1)$ instead of the correct answer $T_3 \simeq -\Omega^{-0.6}(n - 5/7)$. Even the sign of $T_3^Z$ is wrong for $-1 < n < 5/7$ ! It is also interesting to note that on scales where one might want to measure $T_3$, the effective index of the power spectrum is close to $n = -1$. Then Zel'dovich approximation leads to an essentially unskewed PDF, whose shape is mainly governed by its kurtosis (which is also not computed exactly appropriately).

*2.7.3 Approximation of Non-Linear Dynamics* So far, we have used Lagrangian perturbative solutions at the appropriate order to obtain "exact" results, in the regime when such an approach should be applicable. But one may also think of using these solutions as approximation to the real dynamics in the not-so-weakly non-linear regime. Indeed, Zel'dovich [43] idea was to use his ballistic approximation (where gravitational acceleration is ignored), even when the density contrast $\delta$ becomes large. It lead to the development of pancake theory. Of course, as stated by Zel'dovich, "the analytic evaluation of the error is extremely difficult". This lead to many comparisons of Zel'dovich approximation with the exact dynamics, either with simulations (*e.g.*, Doroshkevich *et al.* 1980) or rigorous Eulerian perturbative theory (Grinstein and Wise 1986, Juszkiewicz *et al.* 1993).

---

[4]More precisely, $T_3 \simeq -2\Omega^{-3/5}(1 + \frac{6}{7}\Omega^{-(1/35)-(2/63)})$. If instead of using eq. (26) one uses limited expansions around $\Omega = 1$, *i.e.*, $f_1 \simeq \Omega^{4/7}$ and $f_2 \simeq 2\Omega^{5/9}$, then one obtains a slightly different expression $T_3 \simeq -2\Omega^{-3/5}(1 + \frac{6}{7}\Omega^{-(1/21)})$, as in [6].

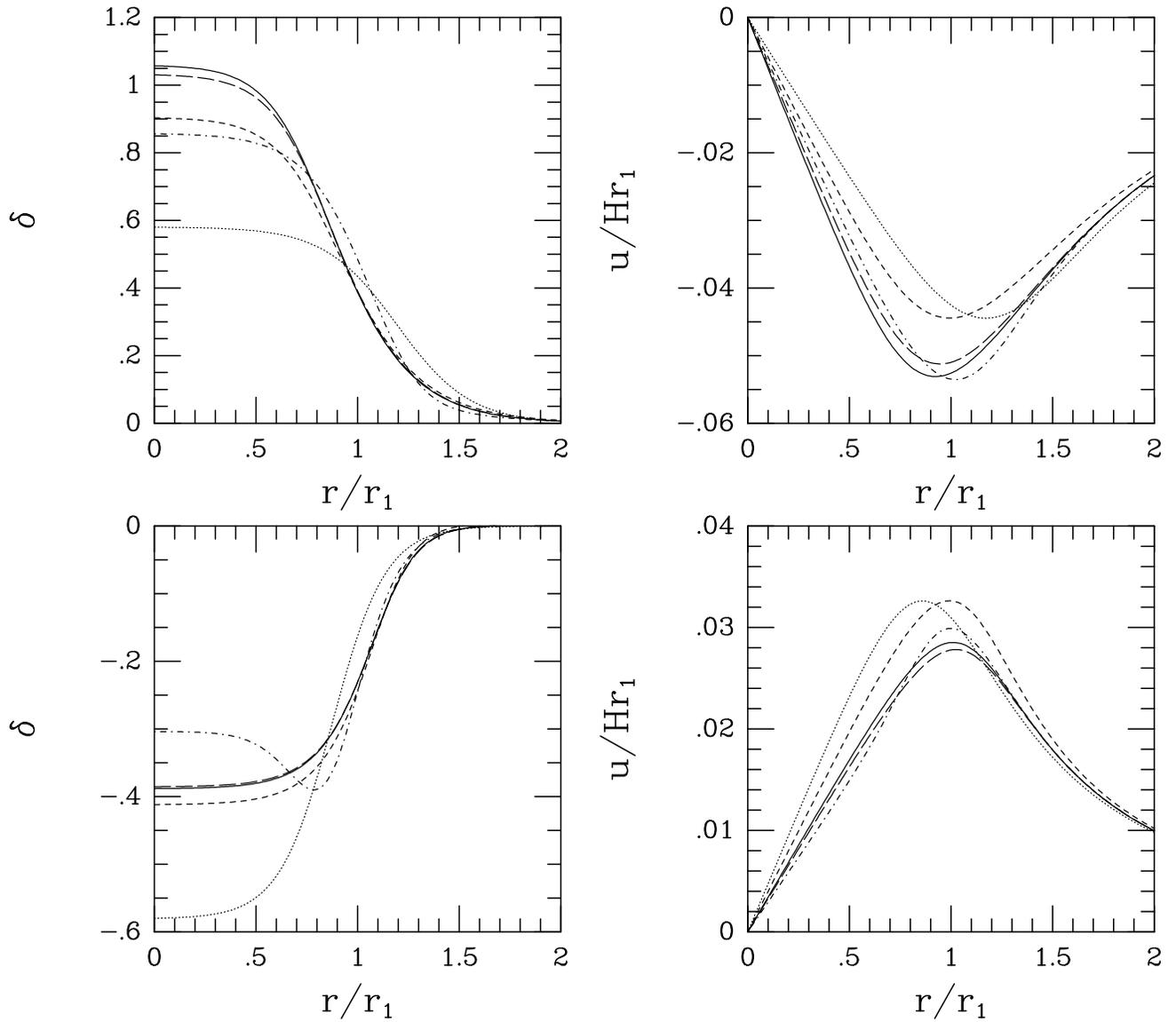

Figure 5: Computed amplitude $\delta$ (left panels) and velocity field (right panels), for a spherically symmetric overdensity (top panels) and an underdensity (bottom panels). The solid line shows the exact result. The long dashes correspond to the second-order Lagrangian approximation, the short dashes to the first order one, *i.e.*, Zel'dovich approximation, the short dashes-dots to the second-order Eulerian approximation, and the dots to Eulerian linear theory. This is for the case $\Omega = 0.1$, but the results do not depend much on this value. Courtesy Bouchet *et al.* 1993 (Ref. [11]).

Despite its limitations, this approximation turns out to be amazingly good, at least when the initial field is smooth enough. Indeed, it is widely used today[5], to the point of being employed to address questions where it is not really appropriate. The success of Zel'dovich approximation brought about many attempts to do better by correcting its shortcomings, and a number of them were discussed in this meeting. The question of how good are those approximations to describe the exact dynamics thus naturally arises. Here we restrict ourselves to rigorous perturbative expansion.

---

[5] For instance to predict the weakly non-linear evolution of the moments of the one point probability distribution function of the density field (Betancort-Rijo 1991; see also Hoffman 1987 for the variance only), or of the distribution itself (Kofman *et al.* 1993, Padmanabhan 1993), or in reconstruction methods to recover the "initial conditions" from present day observations (*e.g.*, Nusser *et al.* 1991, Nusser and Dekel 1992, Gramman 1992, Lachièze-Rey 1993a).

We have compared in [11] the first and second order solutions in Eulerian and Lagrangian perturbation theory to spherically symmetric cases whose evolution is analytically known. For instance, the results of those approximations were checked for the density and the divergence of the velocity field in the spherical top-hat case, when its amplitude is varied. The figure 5 shows the result of another comparison, when the profile is smooth. These lead us to the following approximate ranking (at least for moderate final density contrasts $\sim 1$):

**density contrast :**
Lagrangian second order  $>$ Zeldovich $\gtrsim$
Eulerian second order  $>$ Eulerian linear theory,

**Velocity field :**
Lagrangian second order  $\sim$ Eulerian second order $>$
Zeldovich  $\sim$ Eulerian linear theory.

Here the signs ">" and "$\sim$" mean respectively "more accurate than" and "of comparable accuracy to". For relatively large final density contrasts, the Eulerian approach becomes particularly inefficient, except for the velocity field, for which it tends however to be less accurate than the Lagrangian one. The second order Lagrangian approach gives, for moderate final $\delta$, an excellent approximation of the density contrast and the velocity field. Its seems to be able to reproduce density contrasts as large as ten.

Direct comparisons with numerical simulations (Ref. [11]) appear to confirm that ordering. In essence we find that second-order Lagrangian perturbation theory works well for density constrast up to unity. We expect only little improvements to be brought at higher orders.

## 3  Application to Data

### 3.1  *IRAS* Density Field

Bouchet *et al.* (Refs.[7] & [10]) have measured the count probability distribution function (CPDF) in a series of 10 volume-limited sub-samples of a deep redshift survey of *IRAS* galaxies[6]. Counts were performed in spherical cells, using redshifts as distance estimators. Among other things, they deduced $S_3$ by computing various centered moments of the CPDF. Figure 6 shows on the left an equal weight average in bins of values of $\log \langle \delta^2 \rangle$ ($\overline{\xi_2}$ is just an equivalent notation in the absence of discreteness corrections[7]), the average value $S_3 = 1.5 \pm 0.5$ (dashes), as well as the values inferred from measurements of $Q$ in the non-linear regime from optical data (triangles), from measurements of skewness and variance on the QDOT sample (error bars on left) and the theoretical prediction (solid line on left) from perturbation theory for a power spectrum of index $n = -1.4$ and bias $b_1 = 1$ (se below).

But since Strauss *et al.* (1992a) show that the galaxy densities in cores of clusters determined from *IRAS* galaxies are systematically lower than those determined from optically selected galaxies, they also did counts were galaxies associated with cluster cores were assigned an extra weight corresponding to the ratio of the optical and *IRAS* density estimates. They refer to those counts as the boosted counts. As the right panel of figure 6 shows, it makes a quite noticeable difference. In particular, the average value of $S_3$ is appreciably higher (for $0.1 < \overline{\xi_2} < 10$), namely $3.71 \pm 0.95$, which is more than twice the value found for the unboosted case. This naturally raises the issue of biasing.

---

[6]The sample consists of 5304 galaxies with 60 micron flux density above 1.2 Jy, selected over 87.6% of the sky. The selection criteria for the galaxies are given in Strauss *et al.* (1990) and Fisher (1992), and the data for the brighter half of the sample are given in Strauss *et al.* (1992b). *IRAS* galaxies are a dilute tracer of the galaxian density field (Strauss *et al.* 1992a), with typically 1/3 the number density of galaxies appearing in optically selected samples of comparable depth. Thus one can explore only the low-density limit, but the large volume covered by this sample allows many independent volumes of a given size at a given number density to be probed.

[7]Otherwise, $\langle \delta^2 \rangle = 1/\overline{N} + \overline{\xi_2}$.

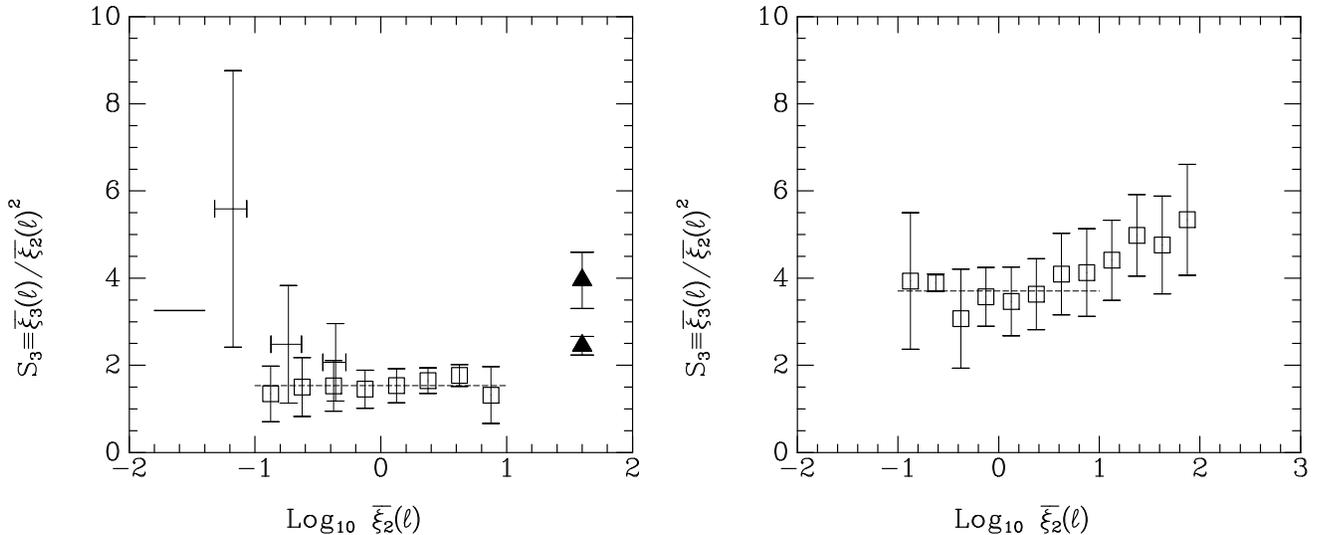

Figure 6: See legend in text. Courtesy Bouchet *et al.* 1993 (Ref. [10])

Indeed, perturbation theory tells us the $S_3$ to expect from the gravitational evolution of gaussian initial conditions. But it is widely believed that the density field traced by galaxies selected in some way (in the optical, or the infrared, *etc.* ), $\delta_g$, may not be identical to the $\delta$ considered so far. Let us suppose, as in [28], that the smoothed galaxy field $\delta_g$ is a local, but non necessarily linear function of $\delta_\ell$, *i.e.*, $\delta_g = B(\delta_\ell)$. Then, by using a Taylor expansion,

$$S_{3g} \equiv \frac{\left\langle \delta_g^3 \right\rangle}{\left\langle \delta_g^2 \right\rangle^2} = \frac{S_3}{b_1} + \frac{3b_2}{b_1^2}, \qquad (29)$$

with $b_1 = B'(0)$ and $b_2 = B''(0)$. This shows that such a biasing preserves the proportionality of the skewness to the square of the variance[8]. One can therefore conclude that the data is compatible with the hypotheses of gaussian primordial fluctuations and local biasing. Furthermore, as mentionned earlier [eq. (4)], a non-gaussian field would yield a term $\propto \sigma$, the proportionality depending of course on the initial value of the skewness $\langle \epsilon^3 \rangle$ (it would also introduce a term involving the initial kurtosis, in the case of a non-linear biasing). One will thus need to specify a particular non-gaussian model (*e.g.*, cosmic strings or textures which are considered in [12] & [3] respectively) to assess whether the data puts a strong constraint on such a model.

3.2 POTENT Velocity Field

The POTENT Velocity Field may be used to check whether relation (11) holds. Since IRAS suggests that the hypothesis of gaussian fluctuations is compatible with the data, one can attempt to use (11) to constraint the value of $\Omega$. Figure 7 shows the measured value of $T_3$ in POTENT samples of various volumes (but at a fixed scale). Small volumes correspond to the best data, while larger volumes offer better statistics. The "best" determination is probably for a volume of radius around 4500 km/s. In order to further assess the effect of sampling, observational errors, finite volume sizes, *etc.* , similar measurements were performed on a series of CDM simulations with 2 different values of $\Omega$ (solid squares correspond to $\Omega = 1$, while solid triangle correspond to $\Omega = 0.3$.

---

[8]Fry & Gaztanaga (1993) have independently derived equation (29), and they have generalized the result in an important way: by expanding $\delta_g = f(\delta)$ in higher order Taylor series, they show that a local biasing function preserves *all* of the moment relations predicted by perturbation theory, in the limit of small fluctuation amplitude.

Taken at face value, the measurement are consistent (within one expected standard deviation) with $\Omega = 1$, while $\Omega = 0.3$ would be excluded at the level of two standard deviations. It should be emphasized, though, that the data analyzed is still preliminary (*i.e.*, it corresponds to today's data, but before it was cleaned, calibrated, and put together properly). Furthermore, the real statistical significance of these "standard deviations" is not really known - the present data is too small to estimate the possible sampling errors due to missing large scale power. One may also worry about the effect of imperfectly correcting for an inhomogeneous Malmquist bias. Our estimate of $\Omega$ is thus more of an illustration of the fact that the *method* should be taken seriously and should have a practical value when larger datasets become available. Future work (in particular concerning the effect of an only partially corrected inhomogeneous Malmquist bias) will tell us how to treat these indications. In any case, it provides an example showing that this method can work.

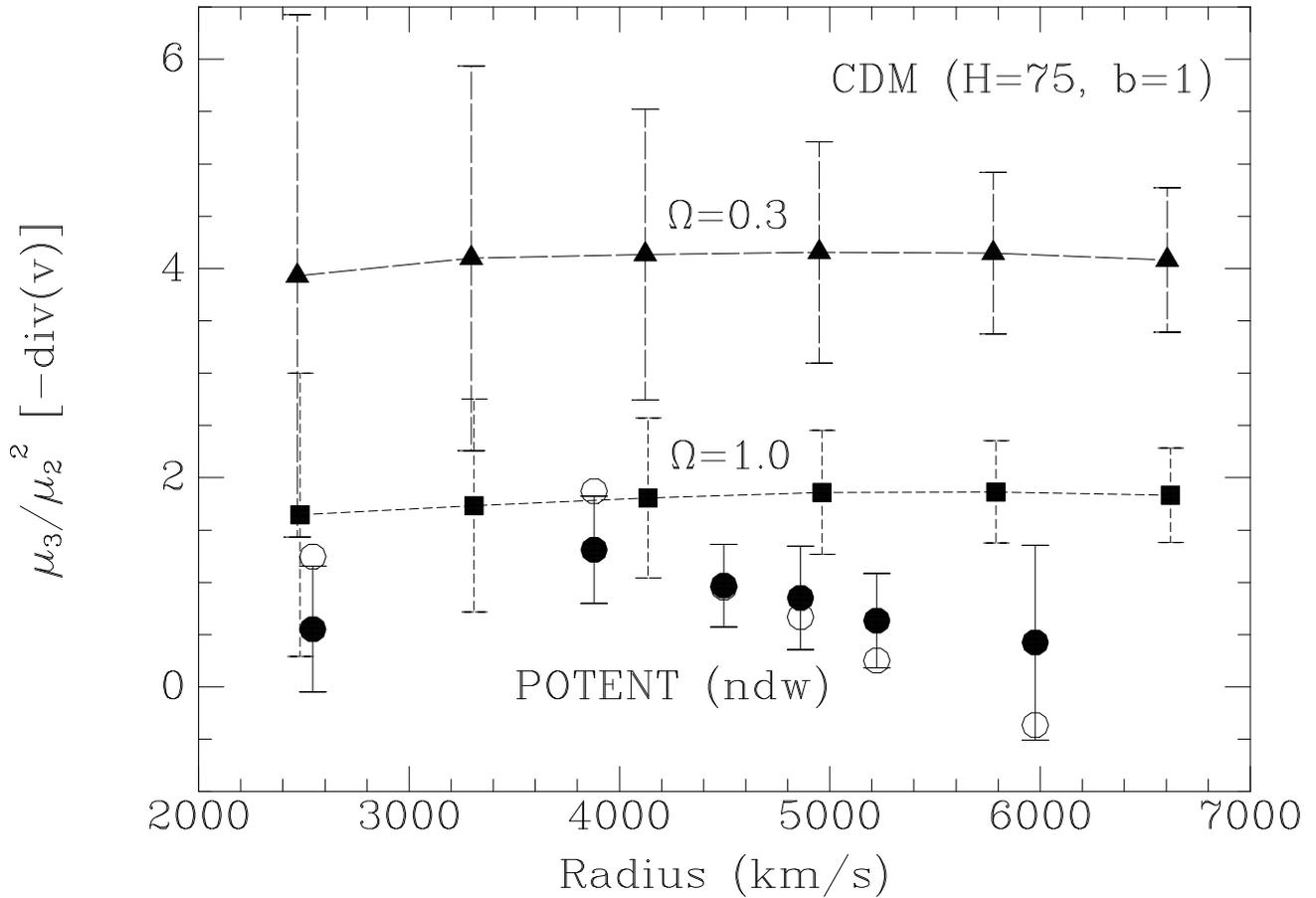

Figure 7: Measurements of $-T_3$ are denoted by solid circles; the attached error bars correspond to a series of measurements made with "fake" data obtained by offsetting the real measurements by some amount given by the expected size of the observational errors, in the standard POTENT analysis manner; the open circles correspond to the raw measurements. See legend in text for further comments. Courtesy Bernardeau *et al.* 1993 (Ref. [6]).

## 4  Summary and Discussion

In this paper, we have described the results of our studies of the relationship between statistics and dynamics of a weakly non-linear, self-gravitating, pressure-free fluid in an expanding Universe. In particular, we have investigated how gravitational instability drives the distribution of density and velocity fluctuations away from the initial state, which we assume gaussian for the most part.

We compared our analytical, perturbative calculations of low-order reduced moments of PDFs of the density and velocity divergence field with N-body simulations. We found excellent agreement over a surprisingly wide dynamical range, all the way to $\sigma \sim 1$ and $\sigma_\theta \sim 1$, when the perturbative series is expected to blow up. Using the Edgeworth expansion, we have calculated the PDFs of $\delta$ and $\theta$. These expansions truncated after the first non-trivial terms turn out to give a good approximation to the real thing provided that $|\nu\sigma| < 1$ and $S_3\sigma < 1$ (and similar requirements for $\sigma_\theta$).

We have investigated the effects of redshift space distortion on the lowest order non trivial cumulant of the PDF of $\delta$ - the skewness $\langle\delta^3\rangle$ - and found that while both skewness and variance are modified by the distortion, the ratio $S_3 \equiv \langle\delta^3\rangle \langle\delta^2\rangle^{-2}$ is little affected, at least in the weakly non-linear regime ($\sigma \lesssim 1$). We have also investigated the possible effects of biasing of the distribution of galaxies with respect to that of mass. We found that when the biasing is a local function of $\delta$, the scaling $\langle\delta^3\rangle \propto \sigma^4$ (expected from gravitationally induced skewness of an initially gaussian field) remains valid for counts of galaxies. These two effects – the lack of redshift distortion and the preservation of the scaling law – can make $S_3$ measurements a powerful tool in distinguishing gaussian initial conditions from strongly non gaussian alternatives, when one expects $\langle\delta^3\rangle \propto \sigma^3$.

The sensitivity of $\langle\theta^3\rangle$ to $\Omega$ can be used to measure the density parameter from the peculiar velocity data alone rather than from heterogeneous data sets (like in the IRAS-POTENT comparison). Such a measurement is also insensitive to whether or not galaxies trace mass. The only requirement is that they do trace the true peculiar velocity field. Estimating $\Omega$ from $\langle\theta^3\rangle$ is somewhat similar to the "reconstruction method" of Nusser & Dekel (1993). An important difference between our two approaches is that Nusser & Dekel use Zel'dovich approximation while we use rigorous perturbation theory. The problem with Zel'dovich approximation is that it fails rather badly in recovering the true numerical values of the reduced moments of the PDFs (in particular those of the smoothed velocity field and/or $\theta$, as was briefly discussed in §2.7.2). Since we showed that the estimate of $\Omega$ depends precisely on these moments, one wonders how accurate can be the resulting estimates of $\Omega$. There is also an observational problem, common to both approaches. The currently available sample may not be deep enough yet (as always seems to be the case in cosmology). The coherence length of the flow appears comparable to the size of the region mapped. Such a sample thus does not appear to be "fair"enough to reliably measure the true distribution of $\theta$, or even its first three moments.

There is clearly more work to be done in the area of perturbation theory, in particular to assess the combined effects of weakly non-gaussian initial conditions, a non-linear bias, for density observations in redshift space. Similarly, further studies are required to check whether perverse effects (finiteness of the sample, inhomogeneous Malmquist bias, *etc.* ) might spoil the simple method using $T_3$ to measure $\Omega$. But, even if these problems taken together turn out to be too formidable to be fully analytically tractable, they are now posed well enough, and the interest of doing so sufficiently well established, that they may be resolved by Monte Carlo analysis.

The ultimate goal of our efforts, presented here, is to use the PDFs of $\delta$ and $\theta$ to test for a generic family of models, based on gravitational instability acting on gaussian initial conditions. So far, we have reliably tested our perturbative calculations with N-body experiments. There are indications of agreement with real data as well. Hopefully, in the near future, new generations of catalogs, like the Digital Sky Survey, will finally tell us whether all we are looking at is just a gaussian random process modified by plain gravity.

**Acknowledgements.** We are grateful to all our collaborators, in particular Piotr Amsterdamski, Jim Bartlett, Francis Bernardeau, Stéphane Colombi, Marc Davis, Avishai Dekel, Eric Hivon, Lars Hernquist, R. Pellat, Michael Strauss, David Weinberg, with whom a part or another of the work reviewed here has been done, and who permitted using some unpublished material.